\documentclass[12pt]{article} 
\usepackage{amsfonts,amsmath,amsxtra}
\usepackage{latexsym}
\usepackage{amssymb}

\def\hybrid{\topmargin 0pt      \oddsidemargin 0pt
        \headheight 0pt \headsep 0pt
        \textwidth 16.5cm
        \textheight 23cm
        \marginparwidth 0.0in
        \parskip 5pt plus 1pt   \jot = 1.5ex}
\catcode`\@=11
\def\marginnote#1{}
\newcount\hour
\newcount\minute
\newtoks\amorpm
\hour=\time\divide\hour by60 \minute=\time{\multiply\hour by60
\global\advance\minute by-\hour}
\edef\standardtime{{\ifnum\hour<12 \global\amorpm={am}%
        \else\global\amorpm={pm}\advance\hour by-12 \fi
        \ifnum\hour=0 \hour=12 \fi
      \number\hour:\ifnum\minute<10 0\fi\number\minute\the\amorpm}}
\edef\militarytime{\number\hour:\ifnum\minute<10 0\fi\number\minute}

\def\draftlabel#1{{\@bsphack\if@filesw {\let\thepage\relax
   \xdef\@gtempa{\write\@auxout{\string
      \newlabel{#1}{{\@currentlabel}{\thepage}}}}}\@gtempa
   \if@nobreak \ifvmode\nobreak\fi\fi\fi\@esphack}
        \gdef\@eqnlabel{#1}}
\def\@eqnlabel{}
\def\@vacuum{}
\def\draftmarginnote#1{\marginpar{\raggedright\scriptsize\tt#1}}

\def\draft{\oddsidemargin -0.1truein
        \def\@oddfoot{\sl preliminary draft \hfil
        \rm\thepage\hfil\sl\today\quad\militarytime}
        \let\@evenfoot\@oddfoot \overfullrule 3pt
        \let\label=\draftlabel
        \let\marginnote=\draftmarginnote
\def\@eqnnum{{\rm (\theequation)}
\rlap{\kern\marginparsep\tt\@eqnlabel}%
\global\let\@eqnlabel\@vacuum}  }

\newfont{\Bbbb}{msbm7 scaled 1\@ptsize00}
\newcommand{\zs}{\raise-1pt\hbox{$\mbox{\Bbbb Z}$}}

scaled 1\@ptsize00

\font\sevenmsa=msam6 
\newfam\msafam
\textfont\msafam=\sevenmsa
\def\hexnumber@#1{\ifnum#1<10 \number#1\else
\ifnum#1=10 A\else\ifnum#1=11 B\else\ifnum#1=12 C\else \ifnum#1=13
D\else\ifnum#1=14 E\else\ifnum#1=15 F\fi\fi\fi\fi\fi\fi\fi}
\def\msa@{\hexnumber@\msafam}
\def\llcorner{\delimiter"4\msa@78\msa@78 }
\def\lrcorner{\delimiter"5\msa@79\msa@79 }
\mathchardef\blacktriangleright="3\msa@49
\mathchardef\blacktriangleleft="3\msa@4A \font\tenmsb=msbm10 scaled
1\@ptsize00
\newfam\msbfam
\textfont\msbfam=\tenmsb \scriptfont\msbfam=\tenmsb

\newdimen\Squaresize \Squaresize=14pt
\newdimen\Thickness \Thickness=0.5pt

\def\Square#1{\hbox{\vrule width \Thickness
   \vbox to \Squaresize{\hrule height \Thickness\vss
      \hbox to \Squaresize{\hss#1\hss}
   \vss\hrule height\Thickness}
\unskip\vrule width \Thickness} \kern-\Thickness}

\def\Vsquare#1{\vbox{\Square{$#1$}}\kern-\Thickness}

\def\numberbysection{\@addtoreset{equation}{section}
        \def\theequation{\thesection.\arabic{equation}}}
\numberbysection

\renewcommand{\theequation}{\thesection.\arabic{equation}}
\def\titlepage{\@restonecolfalse\if@twocolumn\@restonecoltrue\onecolumn
     \else \newpage \fi \thispagestyle{empty}\c@page\z@
        \def\thefootnote{\fnsymbol{footnote}} }

\def\endtitlepage{\if@restonecol\twocolumn \else  \fi
        \def\thefootnote{\arabic{footnote}}
        \setcounter{footnote}{0}}  
\relax

\hybrid
\makeatletter
\newdimen\normalarrayskip            
\newdimen\minarrayskip               
\normalarrayskip\baselineskip \minarrayskip\jot
\newif\ifold             \oldtrue            \def\new{\oldfalse}
\def\arraymode{\ifold\relax\else\displaystyle\fi}
\def\eqnumphantom{\phantom{(\theequation)}} 
\def\@arrayskip{\ifold\baselineskip\z@\lineskip\z@
     \else
     \baselineskip\minarrayskip\lineskip1\baselineskip\fi}

\def\@arrayclassz{\ifcase \@lastchclass \@acolampacol \or
\@ampacol \or \or \or \@addamp \or
   \@acolampacol \or \@firstampfalse \@acol \fi
\edef\@preamble{\@preamble
  \ifcase \@chnum
     \hfil$\relax\arraymode\@sharp$\hfil
     \or $\relax\arraymode\@sharp$\hfil
     \or \hfil$\relax\arraymode\@sharp$\fi}}

\def\@array[#1]#2{\setbox\@arstrutbox=\hbox{\vrule
     height\arraystretch \ht\strutbox
     depth\arraystretch \dp\strutbox
width\z@}\@mkpream{#2}\edef\@preamble{\halign \noexpand\@halignto
\bgroup \tabskip\z@ \@arstrut \@preamble \tabskip\z@ \cr}%
\let\@startpbox\@@startpbox \let\@endpbox\@@endpbox
  \if #1t\vtop \else \if#1b\vbox \else \vcenter \fi\fi
  \bgroup \let\par\relax
  \let\@sharp##\let\protect\relax
  \@arrayskip\@preamble}
\def\eqnarray{\stepcounter{equation}%
              \let\@currentlabel=\theequation
              \global\@eqnswtrue
              \global\@eqcnt\z@
              \tabskip\@centering              
              \let\\=\@eqncr
              $$%
            \halign to \displaywidth  \bgroup
             \eqnumphantom \@eqnsel
      \hskip\@centering                               
    $\displaystyle  \tabskip\z@ {##}$%
    &\global\@eqcnt\@ne \hskip 2\arraycolsep
         $ \displaystyle  \arraymode{##}$\hfil
    &\global\@eqcnt\tw@ \hskip 2\arraycolsep
         $\displaystyle\tabskip\z@{##}$\hfil
         \tabskip\@centering
    &{##}\tabskip\z@\cr}
\makeatother

\newcommand\bqa{\begin{eqnarray}}
\newcommand\eqa{\end{eqnarray}}
\def\be{\begin{eqnarray}\new\begin{array}{cc}}
\def\ee{\end{array}\end{eqnarray}}

\def\beq{\begin{equation}}
\def\eeq{\end{equation}}
\def\bse{\begin{subequations}}                
\def\ese{\end{subequations}}
\def\bp{\begin{pmatrix}}
\def\ep{\end{pmatrix}}

\def\stack#1#2{\raise0.7pt\hbox{$\mathrel{\mathop{#2}\limits^{#1}}$}}
\def\tr{\triangleright}
\def\tl{\triangleleft}
\def\sem{\mathsurround=0pt \raise1pt
\hbox{$\scriptscriptstyle>\!\!$}\:\!\!\tl}
\def\mes{\mathsurround=0pt \tr\!\:\!\raise0.8pt
\hbox{$\scriptscriptstyle\!\!<$}\,}
\def\]{\mathsurround=0pt ]\raise-2pt\hbox{$_\ast$}}

\def\<{\langle}
\def\>{\rangle}

\def\we{\raise-1pt\hbox{$\,\stackrel{\wedge}{,}\,$}}
\def\tr{{\rm tr}\,}

\newcounter{pac}[section]

\newcounter{pacc}[subsection]

0

0

\begin{document}


\vspace{15 mm}
\centerline{\Large \bf Coadjoint Orbits, Cocycles}
\vspace{5 mm}
\centerline{\Large \bf  and} 
\vspace{5 mm}
\centerline{\Large \bf Gravitational Wess-Zumino}
\vspace{8 mm}
\centerline{Anton Alekseev and Samson L. Shatashvili}
\vspace{2 mm}

\vspace{5 mm}
\begin{abstract}

About 30 years ago, in a joint work with L. Faddeev we introduced a geometric action on coadjoint orbits. This action, in particular, gives rise to a path integral formula for characters of the corresponding group $G$. In this paper, we revisit this topic and observe that the geometric action is a 1-cocycle for the loop group $LG$. In the case of $G$ being a central extension, we construct Wess-Zumino (WZ)  type terms and show that the cocycle property of the geometric action gives rise to a Polyakov-Wiegmann (PW)  formula with boundary term given by the 2-cocycle which defines the central extension. In particular, we obtain a PW type formula for the Polyakov's gravitational WZ action. After quantization, this formula leads to an interesting bulk-boundary decoupling phenomenon previously observed in the WZW model. We explain that this decoupling is a general feature of the Wess-Zumino terms obtained from geometric actions, and that in this case the path integral is expressed in terms of the 2-cocycle which defines the central extension.

\vspace{.5cm}

\hspace{6 cm} {\it In memory of our teacher Ludwig Faddeev}

\end{abstract}

\vspace{1cm}


\section{Introduction}

When we were invited to contribute a paper in the Ludwig Faddeev memorial volume, we turned to 
a project that we started together with him more than 30 years ago. And when we started working on the article 
it was a pleasant surprise that many interesting things can be added after so many years. 

Cocycles and geometric actions were the topics of our joint papers with Faddeev \cite{FS1}, 
\cite{FS2}, \cite{AFS}, and this theme continued  in \cite{AS1}, \cite{AS2}, \cite{AS3}. Recent developments  in a wide variety of topics in modern theoretical physics turned out to be connected to those old ideas. To list just a 
few interesting connections, we mention the SYK model \cite{SY}, \cite{K} as studied in \cite{TV}, \cite{MS}, \cite{J}, \cite{MSY}, \cite{EMV}, \cite{MNW} \cite{SW}, \cite{GR} and the vacuum decay in CFT of \cite{PPT}. 

Faddeev advocated  that one should think of quantization as of a path integral in the Hamiltonian formalism
$$
Z=\int \, \mathcal{D}p \mathcal{D}q  \, e^{\frac{i}{\hbar} S}.
$$
Here the action is written as
$$
S=\int \, pdq - H(p,q)dt=\int \, d^{-1} \omega -Hdt,
$$ 
$\omega$ is the symplectic form on phase space and $d^{-1}$ is understood in the sense of Novikov. The integration measure in the path integral is induced by the the Liouville measure on the phase space:
$$
\mathcal{D}p \mathcal{D}q= \prod_t dp(t)dq(t).
$$

In \cite{AFS}, we considered a coadjoint orbit of a compact Lie group $G$ as a phase space. For $G=SU(n), SO(n)$ we chose (following Guillemin-Sternberg \cite{GS}) a  Darboux chart and imposed  boundary conditions on the coordinates $q(t)$ at $t=t_1; t=t_2$ such that the path integral reproduced the correct character formula.

In the present paper, we revisit the construction of geometric actions on coadjoint orbits. It turns out that the geometric action for the group $G$ is a group 1-cocyle for the loop group $LG$. This cocycle property gives rise to an interesting identity for path integrals of the geometric action relating matrix elements before and after the gauge transformation.

We then follow \cite{AS1} and focus on the case when the group $G$ is a central extension 
$$
1 \to S^1 \to G \to G' \to 1
$$
of some other group $G'$ by the circle $S^1$. The main examples are $G$ being the standard central extension of a loop group of a compact connected Lie group and $G$ being the Virasoro-Bott central extension of the diffeomorphism group of the circle. We observe that a linear combination of geometric actions for the three groups ($G$, $G'$ and $S^1$) gives rise to the Wess-Zumino (WZ) type term:
$$
{\rm WZ}=S_G - S_{G'} - S_{S^1}.
$$
 In our examples, this term is given by the action of the WZW model and by the action of the Polyakov gravitational WZ model, respectively.

Using the cocylce property of the geometric action, we prove a Polyakov-Wiegmann (PW) type formula for the WZ terms.\footnote{
When this paper was completed, we learnt that the theory of PW formulas for WZ terms was developed by Aratyn-Nessimov-Pacheva and their collaborators in a series of papers, see {\it e.g} \cite{ANP1, ANP2} } The boundary contribution in this formula is given by the 2-cocylce $\alpha$ which defines the central extension.

We focus our attention on the gravitational Wess-Zumino action. In quantum theory, the corresponding PW formula gives rise to an interesting bulk-boundary decoupling phenomenon where the path integral is completely determined by the classical action with given boundary conditions. This phenomenon was first observed in \cite{BSS} for the WZW model (it is motivated by the background independent open string field theory \cite{W1}, \cite{W2}, \cite{S1}, \cite{S2} and the open/closed string correspondence). We use our formalism to establish the bulk-boundary decoupling for the gravitational WZ action. Our calculation is inspired by the recent studies in \cite{PPT}. Among other things, we observe that the path integral is given by the 2-cocylce $\alpha_2$.

\vskip 0.2cm

{\bf Acknowledgements.} We are grateful to N. Nekrasov, G. Pimentel and A. M. Polyakov for important discussions. We would like to thank F. Valach for useful comments. We are indebted to E. Nissimov and S. Pacheva, former students of L. Faddeev, who attracted our attention to their work on geometric actions and in particular to \cite{ANP1, ANP2}.
The research of A.A. was supported in part by the ERC project MODFLAT, by the grants number 165666, 159581 of the Swiss National Science Foundation and by the National Center of Competence in Research SwissMAP. The research of S.Sh. was supported by Simons Foundation under the program "Targeted Grants to Institutes".

\section{Coadjoint orbits and cocycles}

In this section, we recall the construction of geometric actions on coadjoint orbits of Lie groups (orbit actions), show that they satisfy the 1-cocycle property and compare the Kirillov character formula and the path integral character formula which makes use of an orbit action.

\subsection{Orbit actions and the cocycle property}
Let $G$ be a connected Lie group, $\mathfrak{g}$ be the Lie algebra of $G$ and $\mathfrak{g}^*$ its dual space. The group $G$ acts on $\mathfrak{g}$ by the adjoint action $g: x \to {\rm Ad}_g x$ and on $\mathfrak{g}^*$ by the coadjoint action  $g: \xi \to {\rm Ad}_g^* \xi$. To simplify notation, we pretend that $G, \mathfrak{g}$ and $\mathfrak{g}^*$ are realized as spaces of matrices (this is the case in many examples) and write ${\rm Ad}_g x = g^{-1}xg$ and ${\rm Ad}_g^* \xi = g^{-1} \xi g$. For $\lambda \in \mathfrak{g}^*$ we define its coadjoint orbit
$$
\mathcal{O}_\lambda = \{ \xi= g^{-1}\lambda g; g \in G \}.
$$
There is a canonical projection $\pi_\lambda: G \to \mathcal{O}_\lambda$ given by $\pi_\lambda(g) = g^{-1} \lambda g$. Coadjoint orbits carry canonical Kirillov-Kostant-Souriau (KKS) symplectic forms. It is easy to describe their pull-backs to $G$ under projection $\pi_\lambda$:
$$
\pi_\lambda^* \omega_\lambda = \frac{1}{2} \langle \lambda, [{\rm d}g g^{-1}, {\rm d}g g^{-1}]\rangle = \langle \lambda, ({\rm d}g g^{-1})^2 \rangle = 
 {\rm d} \langle \lambda, {\rm d}g g^{-1} \rangle.
$$
Here ${\rm d}$ is the de Rham differential and ${\rm d}g g^{-1} \in \Omega^1(G, \mathfrak{g})$ is the right-invariant Maurer-Cartan form on $G$. The 2-form $\omega_\lambda$ defines a non-trivial cohomology class in $H^2(\mathcal{O}_\lambda, \mathbb{R})$. The Liouville form $\omega_\lambda^N/N!$, where $N={\rm dim}(\mathcal{O}_\lambda)/2$, is a $G$-invariant volume form on the orbit $\mathcal{O}_\lambda$. 

The interest in the study of coadjoint orbits originates in the Borel-Weil-Bott Theorem and in the Kirillov orbit method which establish a correspondence between irreducible unitary representations of $G$ and coadjoint orbits.  One of the manifestations of this correspondence is the Kirillov character formula \cite{Ki}. We write it in the case of $G$ compact and connected: let $x \in \mathfrak{g}$ and $\exp(x) \in G$ be an element of the Lie algebra and its image under the exponential map, $\lambda$ be the highest weight of the irreducible representation $\sigma_\lambda: G \to {\rm End}(V_\lambda)$ and $\chi_\lambda(g)={\rm Tr}_{V_\lambda} \sigma_\lambda(g)$ be its character. Then, the Kirillov character formula states
\begin{equation} \label{Kirillov_character}
\chi_\lambda(\exp(x)) =  j^{-1/2}(x) \int_{\mathcal{O}_{\lambda+\rho}} \, \frac{\omega_{\lambda+\rho}^N}{N!} \, e^{i\langle \xi, x\rangle} ,
\end{equation}
where $\rho$ is the half sum of positive roots of $G$ and $j(x)$ is the Jacobian of the exponential map $x \mapsto \exp(x)$ with respect to the Lebesgue measure on $\mathfrak{g}$ and the Haar measure on $G$.

The KKS symplectic form allows to define an action of a Hamiltonian system on a coadjoint orbit. Fix $\lambda \in \mathfrak{g}^*$ and consider maps 
$g: I \to G$ for $I$ a 1-dimensional manifold (a segment or a circle). The orbit action is given by the following expression:
\begin{equation}  \label{orbit_action}
S(\lambda, g) =  \int_I \langle \lambda, {\rm d}g g^{-1} \rangle = \int_I \langle \xi, g^{-1}{\rm d}g \rangle,
\end{equation}
where $\xi = g^{-1}\lambda g$. Observe that $S(\lambda, g)$ is a 1-cocycle for the group of paths (or loops if $I=S^1$) $\{ g: I \to G\}$ with point-wise multiplication. That is, $\delta S=0$ where
\begin{equation} \label{1-cocycle}
\delta S(\lambda, g, h)=S(\xi=g^{-1}\lambda g, h) - S(\lambda, gh) + S(\lambda, g).
\end{equation}
This observation plays a key role in the rest of the paper.

\subsection{The path integral character formula}
The main feature of the orbit action is the following path integral identity. Let $A \in \Omega^1(I, \mathfrak{g})$ be a connection on $I$ (all the $G$-bundles on $I$ are trivial). Then, for $I=S^1$ we have the following character formula
\begin{equation} \label{path_character}
\chi_\lambda\left( {\rm Pexp} \int_I A\right) = \int \mathcal{D} g \, e^{i \left( S(\lambda, g) + \int_I \langle \xi, A \rangle \right)} .
\end{equation}
Here ${\rm Pexp} \int_I A$ is the holonomy of $A$ along $I$ (with some base point chosen on the circle).  The path integral is loosely written with the integration measure $\mathcal{D} g$ even though $\mathcal{D} \xi$ would be more appropriate (there is a gauge symmetry preserving $\xi$). In \eqref{path_character}, we are using the coadjoint orbit $\mathcal{O}_\lambda$ but choosing a different regularization of the path integral allows to use the coadjoint orbit $\mathcal{O}_{\lambda +\rho}$ as in the Kirillov character formula. The significant difference of the path integral formula \eqref{path_character} from the finite-dimensional formula \eqref{Kirillov_character} is the absence of the normalization factor $j^{-1/2}(x)$ which is automatically taken into account by the path integral.

In the case of $I=[0, T]$, the path integral defines matrix elements of the representation $\sigma_\lambda$:
\begin{equation} \label{path_segment}
\langle a | \sigma_\lambda\left( {\rm Pexp} \int_I A\right) |b\rangle  = \int_{g(0) \in \mathcal{L}_b, g(T) \in \mathcal{L}_a} \mathcal{D} g \, e^{i \left( S(\lambda, g) + \int_I \langle \xi, A \rangle \right)} .
\end{equation}
Here $\langle a|$ and $|b \rangle$ label two vectors in $V_\lambda$, and $\mathcal{L}_a, \mathcal{L}_b$ are appropriate boundary conditions which correspond to these vectors. For $G=SU(n)$ and $G=SO(n)$ such boundary conditions for elements of the Gelfand-Zeitlin basis were constructed in \cite{AFS}.

Note that under the gauge transformations 
$$
A\mapsto A^h=hAh^{-1} + d hh^{-1}
$$ 
the orbit action transforms as follows:
$$
\begin{array}{lll}
S(\lambda, g) + \int_I \langle g^{-1} \lambda g, A^h\rangle & = & S(\lambda, g) + S(g^{-1} \lambda g, h) + \int_I \langle g^{-1} \lambda g, hAh^{-1} \rangle \\
& = &
S(\lambda, gh) + \int_I \langle (gh)^{-1} \lambda (gh), A \rangle.
\end{array}
$$
Here we have used the cocycle property \eqref{1-cocycle}. This leads to the following transformation property for the 
the path integral on the right hand side of \eqref{path_segment} (for make notation shorter, we drop the boundary conditions):
$$
\begin{array}{lll}
\int \mathcal{D} g \, e^{i \left( S(\lambda, g) + \int_I \langle \xi(g), A^h \rangle \right)} &=& \int \mathcal{D} g \, e^{i \left( S(\lambda, gh) + \int_I \langle (gh)^{-1}\lambda (gh), A \rangle \right)} \\
&=& \int \mathcal{D} v \, e^{i \left( S(\lambda, v) + \int_I \langle \xi(v), A \rangle \right)},
\end{array}
$$
where we have made a substitution $v=gh$. For $I=S^1$, this change of variables doesn't affect the result. For $I=[0, T]$, the boundary conditions $\mathcal{L}_a, \mathcal{L}_b$  get affected and this is represented by the factors $h(0)$ and $h(T)$ in the gauge transformation of the matrix elements of the holonomy:
\begin{equation} \label{holonomy}
\langle a| {\rm Pexp} \int_I A^h |b \rangle = \langle a | h(T) \left( {\rm Pexp} \int_I A\right) h(0)^{-1} | b \rangle.
\end{equation}
In particular, for $A=0$ we obtain the following formula for the path integral:
\begin{equation}   \label{matrix_element}
\int_{g(0) \in \mathcal{L}_b, g(T) \in \mathcal{L}_a} \mathcal{D} g \, e^{i \left( S(\lambda, g) + \int_I \langle \xi, dhh^{-1} \rangle \right)} = 
\langle a | \sigma_\lambda(h(T)) \sigma_\lambda(h(0))^{-1} |b \rangle.
\end{equation}
In what follows, wee will several applications of this formula in Conformal Field Theory.

\section{Central extensions}

In this section, we study in detail orbit actions for central extensions, introduce the notion of a Wess-Zumino action and derive a general Polyakov-Wiegmann (PW) type formula. Our main examples are the Wess-Zumino-Witten (WZW) action which corresponds to the central extensions of  loop groups and the Polyakov gravitational Wess-Zumino action which corresponds to the central extension of the diffeomorphism group of the circle.

\subsection{Orbit actions and the Polyakov-Wiegmann formula}
Let $\widehat{G}$ be a central extension of the group $G$ by the circle $S^1$ defined by an exact sequence of group homomorphisms:
\begin{equation} \label{G_seq}
1 \to S^1 \to \widehat{G} \to G \to 1.
\end{equation}
The corresponding central extension of the Lie algebra $\mathfrak{g}$ is given by the exact sequence of Lie algebra homomorphisms:
\begin{equation}    \label{g_sec}
0 \to \mathbb{R} \to \widehat{\mathfrak{g}} \to \mathfrak{g} \to 0.
\end{equation}

In what follows we will pretend that the $S^1$-bundle $\widehat{G} \to G$ is trivial, and that it admits a section $s: G \to \widehat{G}$. We will need the following objects defined by this section. The group law in $\widehat{G}$ defines a 2-cocycle $\alpha_2: G \times G \to S^1$ such that for $g,h \in G$ one has.
$$
s(g) s(h) =s(gh) e^{i\alpha_2(g,h)}.
$$
The section $s$ induces a splitting $\mathfrak{g} \to \widehat{\mathfrak{g}}$ of the exact sequence \eqref{g_sec} (denoted by the same letter) and  a Lie algebra 2-cocycle $\omega(x,y)$ via the formula
$$
[ s(x), s(y)]_{ \widehat{\mathfrak{g}}}= s([x,y]_\mathfrak{g}) + \omega(x,y).
$$
We introduce the notation
$$
\beta(g)= ds(g) s(g)^{-1} - dgg^{-1} \in \Omega^1(G)
$$
for the 1-form $\beta(g)$  which represents the central part of the Maurer-Cartan form $ds(g) s(g)^{-1}$. 

The splitting $s$ also defines an isomorphism $\widehat{\mathfrak{g}}^* \cong \mathfrak{g}^* \oplus \mathbb{R}$. The coadjoint action of $\widehat{G}$ descends to $G$ since the central circle acts trivially. We will use notation
$$
{\rm Ad}^*_g (\lambda, c) = (g^{-1} \lambda g + c\gamma(g), c),
$$
where $\lambda \in \mathfrak{g}^*, c \in \mathbb{R}$, $g^{-1} \lambda g$ stands for the coadjoint action of $G$ on $\mathfrak{g}^*$  and $\gamma(g) \in \mathfrak{g}^*$ is the contribution of the central extension.

For an element $(\lambda, c) \in \widehat{\mathfrak{g}}^*$, we consider the orbit action for $\widehat{g}=s(g)e^{i\theta}$:
$$
S_{\widehat{G}}((\lambda, c), \widehat{g}) = \int_I \langle (\lambda, c), d\, \widehat{g}\, \widehat{g}^{-1} \rangle =
\int_I ( \langle \lambda, dgg^{-1} \rangle + c(d \theta + \beta(g)) ).
$$
Note that the first term is the orbit action for the group $G$ corresponding to the orbit of $\lambda$ and that the term $S_c(\theta)=c \int_I d\theta$ is the orbit action for the circle $S^1$. The expression 
\begin{equation}    \label{WZ_general}
{\rm WZ}(g)=\int_I \beta(g)
\end{equation}
is independent of $\lambda$ and c and is called the Wess-Zumino action.  To summarize, we have
\begin{equation} \label{orbit_WZ}
S_{\widehat{G}}((\lambda, c), \widehat{g}) = S_{G}(\lambda, g) + S_{S^1}(c, \theta) + c \,  {\rm WZ}(g).
\end{equation}

The orbit actions $S_{\widehat{G}}((\lambda, c), \widehat{g}), S_{G}(\lambda, g)$ and $S_{S^1}(c, \theta)$ satisfy their cocycle conditions. The comparison of these cocycle conditions gives rise to the Polyakov-Wiegmann (PW) type formula of the Wess-Zumino action ${\rm WZ}(g)$. This formula is one of our main results:
\begin{equation}   \label{PW_univ}
{\rm WZ}(g) - {\rm WZ}(gh) + {\rm WZ}(h)= \int_I \langle \gamma(g), dhh^{-1}\rangle + \int_I d\alpha_2(g,h).
\end{equation}
Note that the definitions of ${\rm WZ}(g), \gamma(g)$ and $\alpha_2(g,h)$ depend on the choice of the section $s$. If one chooses another section, all these quantities change so as to preserve the equation \eqref{PW_univ}.  Formula \eqref{PW_univ} (in the case without boundary) was first obtained in \cite{ANP1} using the 1-cocycle property of the expression $\gamma(g)$.

In what follows we will consider two interesting examples of orbit actions for central extensions, of Wess-Zumino terms and their PW formulas.

\subsection{Example: current algebras and WZW action}
Let $G$ be a connected and simply connected semisimple Lie group, $\mathfrak{g}$ be its Lie algebra and $L\mathfrak{g}$ be the space of loops
$x: S^1 \to \mathfrak{g}$. Given an invariant scalar product $B: \mathfrak{g} \times \mathfrak{g} \to \mathbb{R}$, there is  a central extension of $L\mathfrak{g}$ defined by the cocycle
$$
\kappa(\alpha,\beta)= \int_{S^1} \, B(\alpha, d\beta).
$$
As a vector space, the central extension $\widehat{L\mathfrak{g}}$ is a direct sum of the loop algebra $L\mathfrak{g}$ and the real line,  $\widehat{L\mathfrak{g}} = L\mathfrak{g} \oplus \mathbb{R}$. The corresponding group is given by the short exact sequence
$$
1 \to S^1 \to \widehat{LG} \to LG \to 1,
$$
where $LG=\{ g: S^1 \to G\}$ is the loop group of $G$. A convenient way to think of the dual space to $\widehat{L\mathfrak{g}}$ is to identify it with  pairs $(A,c)$, where $c \in \mathbb{R}$ is a real number and $A \in \Omega^1(S^1, \mathfrak{g})$ is a connection on the circle. The pairing with $\widehat{L\mathfrak{g}}$  is given by
$$
\langle (A, c), (x, \mu) \rangle =  \int_{S^1} B(A, x) + c \mu.
$$
The coadjoint action of $\widehat{LG}$ descends to $LG$, and it is given by  gauge transformations
$$
g: (A, c) \mapsto (g^{-1}Ag + c \, g^{-1}\partial_x g, c).
$$
Hence, we identify the map $\gamma(g)=g^{-1} \partial_x g$.
For $c\neq 0$, the coadjoint orbits are in one-to-one correspondence with conjugacy classes in $G$. The correspondence is given by mapping a pair $(c, A)$ to the conjugacy class of the holonomy ${\rm Pexp}\left(\frac{1}{c} \int_{S^1} A\right)$.

The orbit action is naturally defined on $M=I \times S^1$, where $I=[a, b]$ is a segment or a circle $I=S^1$. The variable on $I$ is denoted by $t$ and plays the role of time, the variable on $S^1$ is denoted by $x$ and is interpreted as a space variable.
The orbit action is given by the expression
\begin{equation} 
S((A, c), (g, \theta)) =  \frac{1}{4\pi}\int_{M} \langle a(x), \partial_t gg^{-1} \rangle dtdx + c \int_I d \theta + c\, {\rm WZ}(g)
\end{equation}
where $A=a(x){\rm d}x$.  In this case, the expression ${\rm WZ}(g)$ is the chiral Wess-Zumino-Witten given by the following formula:
\begin{equation} \label{WZW_A}
{\rm WZ}(g)= \frac{1}{4\pi} \int_M \langle \partial_t g g^{-1} , \partial_x g g^{-1} \rangle dtdx + \frac{1}{12\pi} \int_N \langle {\rm d}gg^{-1}, ({\rm d}gg^{-1})^2 \rangle,
\end{equation}
In the case of $M=S^1 \times S^1$, the manifold $N$ is a handlebody with $\partial N = M$, the field $g: M \to G$ extends to $N$ (since $\pi_2(G)$ is trivial) and the second term is defined modulo $2\pi \mathbb{Z}$.  Hence, in this case ${\rm WZ}(g)$ is a multi-valued functional in the sense of Novikov \cite{N}. The coefficients in front of the terms on the right hand side of \eqref{WZW_A} are on the one hand traditional in the literature, and on the other hand they ensure the integrality of the action if the scalar product is the standard invariant scalar product on the Lie algebra $\mathfrak{g}$.

In the case when $M=[0,1] \times S^1$, the action ${\rm WZ}(g)$ should be supplemented with boundary conditions at $t=0, 1$ in order to make the second term in \eqref{WZW_A} well defined.  The ambiguity in the definition of the WZW action arises because the bundle $\widehat{LG} \to LG$ is nontrivial and it does not admit a globally defined section $s$.

The PW formula for the action ${\rm WZ}(g)$ reads
$$
{\rm WZ}(g) - {\rm WZ}(gh) + {\rm WZ}(h) = \frac{1}{4\pi}\int_M \langle g^{-1} \partial_x g, \partial_t hh^{-1} \rangle + \int_I d \alpha_2(g,h).
$$
The first term on the right hand side is given by the classical PW formula of \cite{PW}.  The second term was observed in \cite{BSS} and it is  the 2-cocycle which defines the central extension $\widehat{LG}$. It depends on the choice of a (local) section $s$. For local sections defined  on the neighborhood of the group unit, the universal expressions for $\alpha_2(g,h)$ were classified in \cite{ANXZ}.

\subsection{Example: Virasoro algebra and gravitational WZ}
One more example of an interesting Lie algebra which gives rise to a useful orbit action is the Virasoro algebra. One considers the Lie algebra ${\rm Vect}(S^1)$ of vector fields on the circle with elements of the form $u(x)\partial_x$, where $u(x)$ is a $2\pi$-periodic function. The central extension of ${\rm Vect}(S^1)$  is defined by the Lie algebra Virasoro-Bott 2-cocylce
$$
\omega(u(x) \partial_x, v(x)\partial_x) = \int_{S^1} u(x) v'''(x) dx.
$$
Elements of the dual space are pairs $(B, c)$, where $c$ is a real number (the central charge) and $B=b(x)dx^2$ is a quadratic differential. The coadjoint action of the diffeomorpshim group ${\rm Diff}(S^1)$ is given by 
$$
F: (B=b(x)dx^2, c) \mapsto (B^F= b(F(x)) F'(x)^2 + cS(F)) dx^2, c).
$$ 
Here 
$$
S(F) = \frac{F'''}{F'} - \frac{3 (F'')^2}{2 (F')^2}
$$
is the Schwarzian derivative of $F$. Hence, we identify the element $\gamma(F) = S(F)$.

The orbit action acquires the form
\begin{equation}  \label{S_gr}
S((B, c), (F, \theta)) = \int_M  b(F)F'^2 \, \frac{\dot{F}}{F'}\,  dtdx + c \int_I d\theta + c\, {\rm WZ}_{\rm grav}(F),
\end{equation}
where $b(F)=b(F(x,t))$,  the expression $\dot{F}/F'$ represents the Maurer-Cartan form and the Wess-Zumino term (the gravitational Wess-Zumino action) is given by
\begin{equation}
{\rm WZ}_{\rm grav}= \int_M \,  \frac{ \dot{F}}{F'} \left( \frac{F'''}{F'} - 2 \left( \frac{F''}{F'}\right)^2 \right)  dtdx.
\end{equation}
Note that using the inverse diffeomorphism $f(F(x,t),t)=x$, this expression can be re-written as the Polyakov action for induced 2-dimensional gravity \cite{P}:
\begin{equation} \label{WZ_gr}
\int_M \frac{ \dot{F}}{F'} \left( \frac{F'''}{F'} - 2 \frac{F''^2}{F'^2} \right) dtdx = \int_M \left( \frac{ \dot{f}' f''}{(f')^2} -  \frac{\dot{f} (f'')^2}{(f')^3} \right) dtdy,
\end{equation}
where $y=F(t,x)$.

Our second main result is the PW type formula for the gravitational Wess-Zumino action:
\begin{equation}  \label{PW_grav}
{\rm WZ}_{\rm grav}(F) - {\rm WZ}_{\rm grav}(F \circ G) + {\rm WZ}_{\rm grav}(G) =
\int_M S(F) \, \frac{\dot{g}}{g'} \, dx dt + \int_I d \alpha_2(F,G).
\end{equation}
Here $g(G(x,t), t)=x$ is the inverse diffeomorphism to $G$ and $\alpha_2(F,G)$ is the group Virasoro-Bott 2-cocycle
$$
\alpha_2(F,G) = \int_{S^1} \log(F'(G(x)))' \log(G'(x)) dx.
$$
Note that the structure of the gravitational PW formula \eqref{PW_grav} resembles the standard PW formula. The right hand side is a sum of the bulk term and a boundary term. The boundary term is a cocycle which defines the central extension of the diffeomorphism group. The bulk term is an integral of a product of two factors defined by the two group elements. Note also that there are important new features in the formula \eqref{PW_grav}: the factors in the bulk term are no longer symmetric. One of them is the Schwarzian derivative which is a contribution in the coadjoint action which comes from the central extension. The other one is the Maurer-Cartan form on the diffeomorphism group. In the case of loop groups, these factors resemble each other, and they turn out to be quite different in the case of the diffeomorphism group.

\section{Application: the bulk-boundary decoupling for WZ}

In this section, we apply formula \eqref{matrix_element} to describe certain 2-dimensional path integrals for WZW and gravitational WZ models.
Recall that for $\lambda=0$, the orbit action essentially coincides with the WZ term:
$$
S_{\widehat{G}}((0,c), \widehat{g}) = c \int d\theta + c {\rm WZ}(g).
$$
In what follows, we consistently drop the first term in the action and consider path integrals for the WZ terms\footnote{  Path integrals of the type \eqref{matrix_WZ} (in the case without boundary) were studied in \cite{ANP2}.}:
\begin{equation}    \label{matrix_WZ}
\int_{g(0)\in \mathcal{L}_b, g(T) \in \mathcal{L}_a} \mathcal{D}g e^{ i(c {\rm WZ}(g)) + \int_I \langle \xi, dhh^{-1}\rangle) } =
\langle a|  \sigma_0(h(T)) \sigma_0(h(0))^{-1} |b \rangle.
\end{equation}
We will focus on the situation when the right hand side yields 1 trivializing the path integral.

\subsection{Example: decoupling in WZW}

Here we recall the instance of quantum decoupling observed in the standard WZW model in \cite{BSS} (a slightly different argument was given in  \cite{S3}). To match notation, we replace the real variables $x,t$ with complex variables $z, \bar{z}$.  This implies that the fields $g(z, \bar{z})$ are complex valued (otherwise, the WZW model does not admit solutions of equations of motion).

The boundary condition at one end of the cylinder will be  the vacuum state $|b\rangle = | 0\rangle$ which makes the cylinder into a disc $D$. This state has the following property: it is annihilated by all non-positive components of the holomorphic and anti-holomorphic WZW currents:
$$
J^a_{n} |0\rangle = \bar{J}^a_{n} |0 \rangle =0
$$
for all $n \geq 0$.

The idea is to compute the path integral as a function of values of the field $g$ on the other boundary:
$$
Z(u) = \int_{v|_{\partial D} =u} \mathcal{D}v \,\,e^{i c {\rm WZ}(v)}.
$$
The strategy of \cite{BSS} is to make a substitution $v=g h$, where $h$ is the unique solution of WZW equations of motion with boundary values given by $u$ and $g|_{\partial D}=1$. Using the PW formula, one obtains
$$
Z(u) = e^{i c {\rm WZ}(h)} \, \int_{g|_{\partial D}=1} \mathcal{D}g \, e^{i( c {\rm WZ}(g) + c \int_\Sigma (g^{-1} \partial g, \bar{\partial} h h^{-1}) )} .
$$
Note that the 2-cocycle term $\alpha_2(g,h)=0$ since $g=1$ on the boundary of the disc. The state $\langle a|$ corresponding to the boundary conditions $g|_{\partial D} =1$ is a maximally symmetric Cardy D-brane state (see \cite{C}, and \cite{ASch} for Lagrangian description) which satisfies the boundary conditions
\begin{equation}    \label{Cardy}
\langle a | (J_n-\bar{J}_{-n}) =0
\end{equation}
for all $n \in \mathbb{Z}$.

Classical solutions of equations of motion of the  WZW  model are of the form $h=h_1(\bar{z}) h_2(z)$. Hence,
$$
\bar{\partial} h h^{-1} = \bar{\partial} h_1(\bar{z}) h_1^{-1}(\bar{z})
$$
which is an entire function of $\bar{z}$ on the disc $D$. Note also that $J(z)=g^{-1}\partial g$ is the holomorphic current of the quantum WZW model. Using equation \eqref{matrix_WZ}, we obtain
$$
\int_{g|_{\partial D}=1} \mathcal{D}g e^{ i(c {\rm WZ}(g)) + \int_I \langle \xi, dhh^{-1}\rangle) } =\langle a | \sigma_0(h_1(\bar{z})|_{|z|=1} ) | 0 \rangle.
$$
Here the contribution of $h(0)$ at the center of the disc disappears because of the invariance of the vacuum state $|0\rangle$.
 The group element $h_1(\bar{z})$ is of the form 
 $$
 h_1(\bar{z})=\exp(\sum_{n < 0 ,a} c_{n,a} J^a_n),
 $$ 
 where $n$'s are all negative (non-negative $n$'s correspond to a singularity at $\bar{z}=0$). This yields
$$
\langle a | \sigma_0(\exp(\sum_{n,a} c_{n< 0,a} J^a_{n}) | 0 \rangle = \langle a | \sigma_0(\exp(\sum_{n< 0,a} c_{n,a} \bar{J}^a_{-n}) | 0 \rangle = \langle a| 0\rangle =1,
$$
where we have used the properties of the D-brane state $\langle a|$ and of the vacuum state $|0\rangle$.

In fact,  the argument above applies to any maximally symmetric Cardy state $\langle a|$ which satisfies equation \eqref{Cardy}. For all such states, the quantum decoupling phenomenon occurs. However, the boundary conditions for the field $v=gh$ turn out to be rather involved:
$$
(h(v^{-1} \partial v) h - \partial h h^{-1} - \bar{\partial}vv^{-1} + v h^{-1} \bar{\partial} hv^{-1})|_{\partial D} =0.
$$
Another observation is that the 2-cocycle $\alpha_2(g,h)$ is in general non-vanishing and it contributes to the answer.

In the original setup, we can use the PW formula one more time to we obtain (following \cite{BSS}) a compact and beautiful answer:
\begin{equation}   \label{WZW_answer}
Z(u) = e^{ic {\rm WZ}(h)} = e^{ic \alpha_2(h_1|_{\partial D}, h_2|_{\partial D})}.
\end{equation}
That is, the partition function is completely described in terms of the Riemann-Hilbert decomposition $h=h_1(\bar{z})h_2(z)$ and of the 2-cocycle $\alpha_2$.

\subsection{Decoupling in gravitational WZ}

In this section we apply the example of the WZW calculation of the previous section to the case of the gravitational WZ action.
As before, we impose a vacuum state $|b\rangle = |0 \rangle$ at one end of the cylinder making it into a disc $D$. Recall that the vacuum state is annihilated by Virasoro generators
$$
L_n |0\rangle = \bar{L}_n |0 \rangle =0
$$
for $n\geq -1$. In order to address the decoupling phenomenon, we consider the path integral of the type \eqref{matrix_WZ} for gravitational WZ action:
\begin{equation}  \label{path_int_source}
I(H)=\int_{G|_{\partial D} \in \mathcal{L}} \mathcal{D} G \, e^{i(c {\rm WZ}_{\rm grav}(G) + \int_D S(G) \frac{\bar{\partial} h}{\partial h} d^2z)},
\end{equation}
where $H(z, \bar{z})$ is a classical solution of equations of motion of the action ${\rm WZ}_{\rm grav}$, $h(H(z,\bar{z}), \bar{z})=z$ is the inverse diffeomorphism and $\mathcal{L}$ stands for a conformal boundary condition $\langle a|$ which satisfies the equation
$$
\langle a| (L_n -\bar{L}_{-n}) =0
$$
for all $n$.  In particular, we assume $G|_{\partial D}(z)=z$. 

Solutions of equations of motion $H(z, \bar{z})$ are given by formula
$$
H(z, \bar{z})=\frac{A(\bar{z}) u(z) +B(\bar{z})}{C(\bar{z}) u(z) + D(\bar{z})},
$$
where $AD-BC=1$. Hence, it can be represented as $H=H_2 \circ H_1$, where 
\begin{equation}
\label{RHil}
H_1: z \mapsto u(z), \hskip 0.3cm H_2: z \mapsto \frac{A(\bar{z}) z +B(\bar{z})}{C(\bar{z}) z + D(\bar{z})} .
\end{equation}
Recall (see \cite{AS1}) that the combination $\bar{\partial}h/\partial h$ admits the following presentation
$$
\frac{\bar{\partial}h}{\partial h} = z^2 j_+(\bar{z})  - 2 z j_0(\bar{z}) + j_-(\bar{z}),
$$
where $j_{+,0,-}$ are currents discovered in \cite{P} (they are functions of the parameters $A,B,C,D$ and of their derivatives).

Equation \eqref{matrix_WZ} suggests the following answer for the path integral $I(H)$:
$$
I(H) = \langle a| \sigma_0(H_2) \sigma_0(H_1) |0\rangle .
$$
Note that $H_2$ is a M\"obius transformation generated by $L_{1,0,-1}$. Using the property of the conformal invariant state $\langle a|$, we transform it into a M\"obius transformation generated by $\bar{L}_{1,0,-1}$ which then preserves the vacuum $|0\rangle$. Hence, we have
$$
I(H) = \langle a|  \sigma_0(H_1) |0\rangle .
$$
Similar to the arguments in the previous section, we write 
$$
H_1=\exp(\sum_{n\leq 0} c_n L_n)
$$ 
and compute
$$
I(H)=\langle a| \sigma_0(\exp(\sum_{n\leq 0} L_n)) |0\rangle =
\langle a| \sigma_0(\exp(\sum_{n\leq 0} \bar{L}_{-n})) |0\rangle =\langle a| 0\rangle =1.
$$
Here we have used  conformal invariance of the state $\langle a|$ and the properties of the vacuum.

We are now in the position to compute a more general path integral for the gravitational WZ action:
$$
Z= \int_{V|_{\partial D} \in \mathcal{L}'} \mathcal{D} V e^{ic {\rm WZ}_{\rm grav}(V)} = e^{ic {\rm WZ}_{\rm grav}(H)} I(H) =e^{ic {\rm WZ}_{\rm grav}(H)}.
$$
Here $V = G \circ H$, where $G$ satisfies boundary conditions $\mathcal{L}$ and $H$ is a solution of equations of motion for ${\rm WZ}_{\rm grav}$. Note that the Virasoro-Bott cocycle vanishes: $\alpha_2(G,H)=0$ since $\log(\partial G)=\log(1)=0$ on the boundary of the disc. At the moment, the description of the boundary conditions $\mathcal{L}'$ is implicit. Giving a more explicit description, similar to the WZW case, remains an interesting question (see also \cite{S3}).

Using the gravitational PW formula, we can further simplify the answer for the partition function $Z$:
$$
Z=e^{ic {\rm WZ}_{grav}(H) } = e^{ic \alpha_2(H_1|_{\partial D}, H_2|_{\partial D})}.
$$ 
This is an analogue of equation \eqref{WZW_answer} in the WZW case. We observe that the partition function is completely described in terms of the decomposition \eqref{RHil}. We interpret this decomposition as a gravitational analogue of the Riemann-Hilbert problem.

\vskip 1cm

\noindent {\small {\bf A.A}:  \hspace{.2 cm}  {\sl Department of Mathematics, University of Geneva, Geneva, Switzerland \hspace{8 cm}\,}\\
\hphantom{xxxxxx} {\it E-mail address}: {\tt Anton.Alekseev@unige.ch}}

\noindent{\small {\bf S.Sh.}: {\sl  The Hamilton Mathematics Institute and the School of Mathematics, \hspace{8 cm}\,
\hphantom{xxxx}   \hspace{3 mm} Trinity College Dublin, Dublin 2, Ireland
\hspace{8 cm}\,
\hphantom{xxxx}   \hspace{3 mm} 
Simons Center for Geometry and Physics, Stony Brook, USA 
 \hspace{8 cm}\,
\hphantom{xxxx}   \hspace{3 mm} 
Chair Israel Gelfand, IHES, Bures-sur-Yvette, France
\vspace{1 mm}
 \hspace{8 cm}\,
\hphantom{xxxx}   \hspace{3 mm} 
\vspace{1 mm}
On leave of absence from: \hspace{16 cm}\,
\hphantom{xxxx}   \hspace{3 mm}  Institute for Information Transmission Problems, Moscow, Russia
 }\\
\hphantom{xxxxxx} {\it E-mail address}: {\tt samson@maths.tcd.ie}}

\end{document}

\end{document}